\pdfoutput=1	

\documentclass[10pt]{iopart}

\usepackage{hyperref}
\usepackage{graphicx}

\usepackage{rotating}

\begin{document}


\newcommand{\ub}[1]{\underline{\textbf{#1}}}

\title{Generalized refraction using lenslet arrays}

\author{Alasdair C.\ Hamilton and Johannes Courtial}
\ead{a.c.hamilton@physics.gla.ac.uk}
\address{Department of Physics and Astronomy, Faculty of Physical Sciences, University of Glasgow, Glasgow G12~8QQ, United~Kingdom}

\date{\today}

\begin{abstract}
We have recently started to investigate 2D arrays of confocal lens pairs.
Miniaturization of the lens pairs can make the array behave ray-optically like a homogeneous medium.
Here we generalize the geometry of the lens pairs.
These generalisations include a sideways shift of the lens centres and a change in the orientation of both lenses in each pair.
We investigate the basic ray optics of the resulting arrays, and illustrate these with movies rendered using ray-tracing software.
We suggest that confocal lenslet arrays could be used to realize ray-optically some recent metamaterials concepts such as the coordinate-transform design paradigm.
\end{abstract}

\submitto{\JOA}

\pacs{
01.50.Wg, 
42.15.-i, 
42.70.-a
}

\section{Introduction}

The invention and experimental realization of metamaterials with highly unusual optical properties \cite{Sievenpiper-et-al-1996,Pendry-et-al-1996,Pendry-et-al-1998} such as negative refraction and resonant amplification of evanescent waves has opened up one of the most exciting current research areas in optics. Metamaterials have led to many new concepts, for example superlenses that image not only travelling waves but also evanescent waves \cite{Pendry-2000}, and a new optical-design paradigm based on coordinate transforms~\cite{Leonhardt-2006,Pendry-et-al-2006}.

The manufacture of metamaterials is still exceedingly difficult \cite{Boltasseva-Shalaev-2008}.
Even the simplest metamaterials for near-visible wavelengths (e.g.\ \cite{Smolyaninov-et-al-2007}) are currently not very big (interference lithography can structure areas of order 1cm$^2$ \cite{Boltasseva-Shalaev-2008}) and difficult to manufacture.
Metamaterials are only now beginning to work over a broad band of frequencies \cite{Liu-et-al-2009}.
Our work is motivated by the idea of using standard optical components -- lenslet (or microlens) arrays -- to enable many (but not all) of the unusual ray-optical properties of metamaterials to be ``experienced'' by approximating these metamaterial properties ray-optically on an everyday scale ($\sim$10cm) and across the entire visible spectrum.
Perhaps most excitingly, our ray-optical materials enable light-ray direction changes without wave-optical analog \cite{Hamilton-Courtial-2009}.
Their realization could therefore open up entirely new concepts.

Like with metamaterials, our starting point is negative refraction, specifically the recent realization \cite{Courtial-2008a} that a sheet formed by two identical, parallel, lenslet arrays (LAs) with a common focal plane (so the lenslet arrays are confocal) \cite{Stevens-Harvey-2002,Okano-Arai-2002} acts ray-optically like the interface between optical media with opposite refractive indices.
(Note that wave-optically such a sheet does \emph{not} act like the interface between optical media with opposite refractive indices.  There is, for example, no negative group velocity \cite{Veselago-1968} and no amplification of evanescent waves required for the sub-wavelength imaging properties of superlenses \cite{Pendry-2000} and hyperlenses \cite{Liu-et-al-2007,Smolyaninov-et-al-2007}.)
A sheet of two confocal LAs (CLAs) with different focal lengths approximates the interface between optical media with a refractive-index ratio given by minus the focal-length ratio \cite{Courtial-2008a}.
The refractive-index ratio can be positive or negative. 
This holds only for the part of the light that passes through corresponding lenslets \cite{Courtial-2008b}.
CLAs can be designed such that other light is absorbed; if it is not absorbed, it leads to ``ghost images'' \cite{Courtial-2008b}.

Here we generalize CLAs.
Simple modifications (such as a sideways displacement of the centres of corresponding lenslets in the two LAs) offer additional design parameters that will enable unprecedented control over light rays, while retaining the important property of the structure acting macroscopically like a homogeneous material.
Generalized CLAs can then do significantly more than simply act like the interface between different refractive indices, and this paper provides a concise atlas of distorted views through different individual CLAs.

We speculate that generalized CLAs can offer control over light-ray propagation similar to that afforded by metamaterials.
This could enable, for example, a ray-optical approximation of meta-material concepts such as the coordinate-transform design paradigm \cite{Pendry-et-al-2006}.
Because of this and other analogies with metamaterials we call these generalized CLAs \ub{meta}ma\ub{t}erials f\ub{o}r ra\ub{y}s (METATOYs)~\cite{Hamilton-Courtial-2009}.



\section{Generalization of confocal lenslet arrays}

A CLA sheet consists of two LAs, one with focal length $f_1$, the other with focal length $f_2$, arranged so that they share a common focal plane.

CLAs with $f_1=f_2$ have long been used.
The first application was 3D imaging, initially for ``integral photography'' \cite{Lippmann-1908}, more recently for 3D displays (see Ref.\ \cite{Napoli-et-al-2008} and references therein).
Related to imaging is the moir\'{e} magnifier \cite{Hutley-et-al-1994}, which is based on ``mis-aligned'' CLAs. 
Without making any connection with metamaterials, what can be argued to be the analogs of a superlens \cite{Pendry-2000,Fang-et-al-2005} and a hyperlens \cite{Jacob-et-al-2006,Smolyaninov-et-al-2007,Liu-et-al-2007} have already been built using CLAs \cite{Stevens-Harvey-2002,Okano-Arai-2002,Volkel-et-al-2003,Duparre-et-al-2005}.
However, these do not offer the amplification of evanescent waves and corresponding sub-wavelength-resolution-imaging properties afforded by their metamaterial counterparts, and they are therefore arguably much less interesting.
Nevertheless, all of these are examples of the possibilities afforded by CLAs.

An important property of the CLAs we consider here is that the change of ray direction is independent of the precise point where a light ray hits the array.
If the offset of the light rays on transmission through the CLAs is small (as is the case if the individual lenslets are small \cite{Courtial-2008a}), CLAs therefore have a homogeneous appearance.
This property can also be realized with arrays of prisms \cite{Courtial-Nelson-2008,Hamilton-et-al-2008}.
In CLAs it is due to the following simple argument.
Two parallel light rays, after hitting the first lenslet, are focussed so that they pass through the same point of the lenslet's back focal plane.
But as this plane is also the second lenslet's front focal plane, the two light rays are parallel again after passing through the second lenslet.
This is not normally the case, like for example in a single LA or non-confocal pairs of LAs.

\begin{figure}
\begin{center} \includegraphics{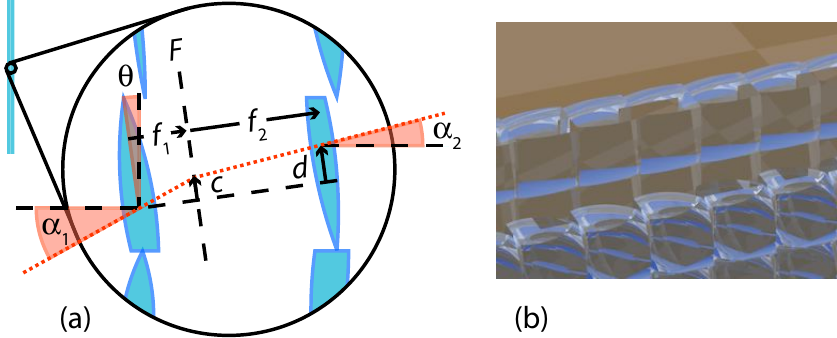} \end{center}
\caption{\label{lens-figure}Geometry and ray optics of generalized confocal lenslet arrays (CLAs).
(a)~Schematic of two-dimensional generalized CLAs.
An individual lens (focal length $f_1$) in the left lenslet array focusses all light rays that impinge on it with the same angle of incidence $\alpha_1$ in a point in the focal plane $F$.
These rays leave the corresponding lens (focal length $f_2$) in the right array with an exit angle $\alpha_2$.
The light rays that pass through the centre of each lens are shown as a dotted line.
Both lenses, and with them their common focal plane, $F$, are inclined with respect to the plane of the CLAs by an angle $\theta$.
In addition, the optical axes of the lenses are displaced by a distance $d$.
(b)~Ray-tracing image of a small part of three-dimensional generalized CLAs.
The picture shows the top edge of the two lenslet arrays, rendered as transparent with a blue tinge, in front of a light-brown/dark-brown checker-board floor.
Each lenslet array consists of identical lenslets, but the lenslets that make up the two arrays are different.
This image was rendered using the ray-tracing software POV-Ray~\cite{POV-Ray}.
}
\end{figure}

To be able to realize metamaterials concepts that generalize refraction, CLAs need to be generalized.
The following generalisations keep corresponding lenslets in the two arrays confocal and therefore the appearance of the CLAs homogeneous, while introducing additional parameters that affect the way the light-ray direction is changed:
\begin{itemize}
\item the centres of corresponding lenslets can be offset with respect to each other;
\item each pair of corresponding lenslets, and with them their common focal plane, can be inclined;
\item the lenslets can be made elliptical, whereby the focal lengths in the directions of the axes of the ellipses of corresponding lenslets have to add up to the same value, namely the separation between the lenses.
\end{itemize}
Fig.\ \ref{lens-figure} shows examples of such generalized CLAs.
Manufacturing such generalized CLAs will require a significant extension of current techniques for manufacturing complex microlens arrays \cite{Hoopman-1995,Duparre-et-al-2005a}.

From now on, we will usually drop the adjective ``generalized'' and simply refer to CLAs.
In the following sections we discuss CLAs in two and three dimensions.

\section{Two-dimensional case: cylindrical CLAs\label{2D-section}}

It is instructive to start by considering the two-dimensional (2D) case.
This case also describes transmission through cylindrical CLAs (more precisely, confocal arrays of cylindrical lenslets), specifically the projection into a plane normal to the cylindrical-lenslet axes. 

Figure \ref{lens-figure}(a) shows a cross-section through CLAs in two dimensions.
The two central lenslets have focal lengths $f_1$ and $f_2$, and they are separated by the sum of their focal lengths, $f_1+f_2$.
The two lenslets' optical axes are parallel; that of the second lenslet is shifted with respect to that of the first by a distance $d$.
In addition, both lenslets, and with them their common focal plane, $F$, are inclined with respect to the vertical (the plane of the CLA sheet) by an angle $\theta$.
This means the optical axes are inclined by the same angle $\theta$ with respect to the sheet normal.

Light rays hitting the left lenslet at an angle $\alpha_1$ with respect to the sheet normal (the horizontal in Fig.\ \ref{lens-figure}(a)) have an angle $(\alpha_1 - \theta)$ with respect to the lenslet's optical axis.
They therefore pass through the common focal plane a distance
\begin{equation}
c = f_1 \tan (\alpha_1-\theta)
\end{equation}
from the left lenslet's optical axis.
Provided they then pass through the corresponding lenslet on the right, such rays leave at an angle $(\alpha_2-\theta)$ with respect to the right lenslet's optical axis such that
\begin{equation}
\tan (\alpha_2-\theta) = (d-c)/f_2.
\end{equation}
$\alpha_2$ is the angle of refraction: the angle with respect to the sheet normal at which the rays leave the sheet.
Eliminating $c$ from these equations gives
\begin{equation}
f_1 \tan (\alpha_1-\theta) = d - f_2 \tan (\alpha_2-\theta).
\end{equation}
With the dimensionless quantities
\begin{equation}
\delta = \frac{d}{f_1}, \label{delta-equation}
\end{equation}
and
\begin{equation}
\eta = - \frac{f_2}{f_1}, \label{eta-equation}
\end{equation}
this becomes
\begin{equation}
\tan (\alpha_1-\theta) = \delta + \eta \tan (\alpha_2-\theta).
\label{tan-equation}
\end{equation}
Equation (\ref{tan-equation}) is one of the key results of this paper.
It describes a generalized ``law of refraction'' for CLAs in two dimensions, or cylindrical CLAs.
In the following section we generalize this to the three-dimensional case.

\section{Three-dimensional case: elliptical CLAs}


We start our generalisation of the 3D case with two confocal spherical lenses with a common optical axis, the $z$ axis.
Such a pair of confocal lenses has the important property that the direction of the refracted ray depends only on the direction of the incident ray and is independent of the exact position where the ray hits the first lens \cite{Courtial-2008a}.
In this section we discuss generalisations of the lens pair that retain this property, and we consider the two lenses to be corresponding lenslets in CLAs.
Fig.\ \ref{lens-figure}(b) shows a ray-tracing image of such generalized CLAs.
It was created using the free, open-source, ray-tracing program POV-Ray \cite{POV-Ray}, which we also used throughout the remainder of this section to illustrate the view through examples of generalized CLAs.

As the first step of our generalisation we replace the spherical lenses with elliptical (or astigmatic) lenses.
We consider the case of elliptical lenses whose major and minor axes respectively point in the $x$ and $y$ directions or vice versa.
(The restriction of the major and minor axes to the $x$ and $y$ directions will be generalized later.)
The effect of an elliptical lens on a light beam can be understood in the following, idealized, way\footnote{Note that the standard geometrical treatment of the effect of spherical lenses on light rays, which we have used in Ref.\ \cite{Courtial-2008a} and in the previous section to derive Eqn (\ref{tan-equation}), is also idealized.}.
An elliptical lens is approximately equivalent to two orthogonal cylindrical lenses
aligned with the elliptical lens's major and minor axes.
We consider the effect of the elliptical lens on a transmitted ray bundle in terms of the orthographic projections into the $x$-$z$ and $y$-$z$ planes.
In each projection we only consider the effect of the cylindrical lens whose corresponding cylinder axis is normal to the projection plane, that is, the lens which has curved surfaces in the projection; we ignore the effect of the other cylindrical lens.
The lens then has two different focal lengths in the two projections, $f_x$ in the $x$-$z$ projection and $f_y$ in the $y$-$z$ projection \cite{OShea-1985-cylindrical-lenses}.


\begin{figure}
\centering \includegraphics{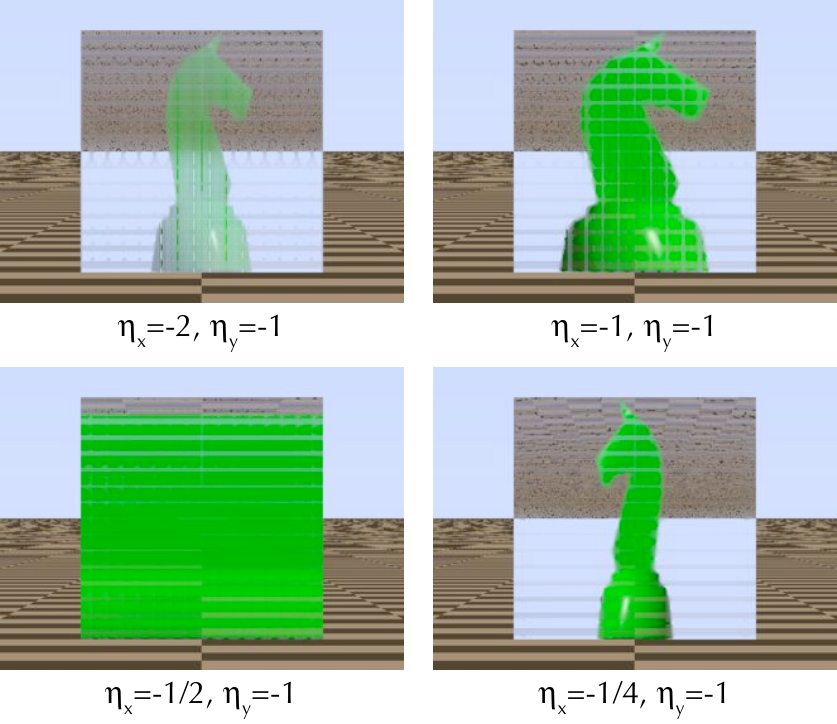}
\caption{\label{vary-etaX-figure}
View through confocal lenslet arrays with different values of $\eta_x$.
Each frame shows the simulated view of a chess piece seen through CLAs.
The CLAs are positioned in the $z=0$ plane, the chess piece is positioned at $z_k = +10$, the camera is positioned at $z_c = -20$.
In this figure (and all subsequent figures), all lengths are given in units of the floor-tile length.
The CLAs consist of $200 \times 200$ confocal lenslet pairs, covering a square area of side length 1.
The separation between corresponding lenslets is 0.05; all lenslets are convex-convex with a centre thickness of 0.0005.
(We restrict ourselves throughout this paper to simulating convex-convex elliptical lenslets, which we describe as the intersection between two similar, but displaced, ellipsoids.)
All views were simulated with POV-Ray \cite{POV-Ray}.
In subsequent figures we use the case $\eta_x = -1/4$, $\eta_y = -1$ as the starting position for further generalisation.}
\end{figure}

To retain the position independence of the ray-direction change, the two lenses have to be confocal in both projections.
In other words, the focal lengths in the $x$-$z$ projection of the first and second lens have to add up to the lens separation, and the same has to be true in the $y$-$z$ projection, so
\begin{equation}
f_{x1} + f_{x2} = f_{y1} + f_{y2}.
\end{equation}
In analogy to the previous definition, we describe the focal-length ratios in the $x$-$z$ and $y$-$z$ projections by the dimensionless quantities
\begin{equation}
\eta_x = - \frac{f_{x2}}{f_{x1}}, \quad \eta_y = - \frac{f_{y2}}{f_{y1}}.
\end{equation}
Fig.\ \ref{vary-etaX-figure} shows a chess piece as seen through different CLAs, all with $\eta_y = -1$ but with different values of $\eta_x$.
Most of the visible part of the chess piece (everything apart from the plinth) is positioned in a plane parallel to the CLAs and appears squashed or stretched in the horizontal ($x$) direction.

Fig.\ \ref{vary-etaX-figure} and all subsequent figures are calculated for CLAs comprising pairs of confocal lenslet pairs with a relatively small (1:100) ratio between lenslet diameter and distance between the lenslets.
This ratio allows all lenslets to be relatively thin and used close to their optical axis, which leads to a respectable image quality.
On the other hand, it severely limits the field of view of the CLAs (approximately to a cone with a cone angle of order $\tan^{-1}(1/100)$).
Consequently, all figures are calculated as if the scene was photographed with a lens with a long focal length, and correspondingly small field of view.
More careful lens design should enable a significant increase in the CLA's field of view without compromising optical quality.

\begin{figure}
\centering \includegraphics{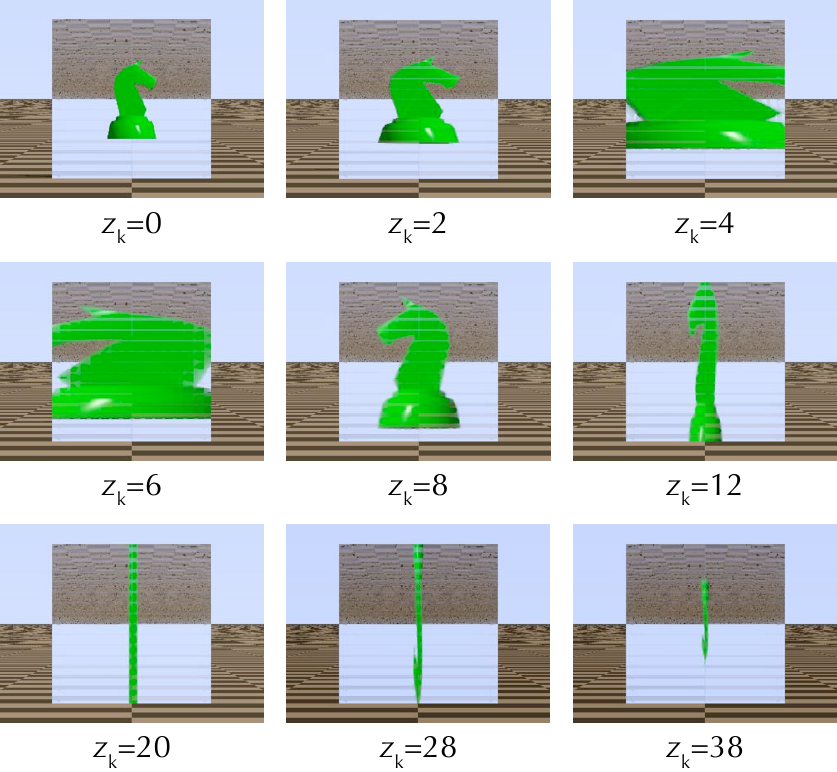}
\caption{\label{vary-knightZ-figure}
Dependence of the apparent stretching of an object seen through CLAs on the distance between the CLAs and the object.
Here the object is a chess piece placed at various distances $z_k$ behind the CLAs.
The frames shown here are taken from a movie showing the view through the CLAs as the chess piece moves between $z_k = -2$ to $z_k = 38$.
The movie is contained in the supporting online material (MPEG-4, 184kB).
This simulation was performed with $\eta_x=-1/4$ and all other parameters like in Fig.\ \ref{vary-etaX-figure}.}
\end{figure}

Fig.\ \ref{vary-knightZ-figure} demonstrates that the squashing and stretching shown in Fig.\ \ref{vary-etaX-figure} depends on the distance between the CLAs and the chess piece.
In the particular configuration for which Fig.\ \ref{vary-knightZ-figure} was calculated, this apparent stretching happens as follows.
As the chess piece is moved further and further away from the CLAs, it first appears stretched in the $x$ direction, whereby the $x$ stretch factor changes sign.
In fact, this sign change occurs at $z_k = 5$, where the $x$ stretch factor diverges:  it becomes $+\infty$ as $z_k = 5$ is approached from below, and $-\infty$ as it is approached from above.
Subsequently the chess piece appears very noticeably stretched in the $y$ direction, whereby the $y$ stretch factor also changes sign (at $z_k = 20$).

\begin{figure}
\centering \includegraphics{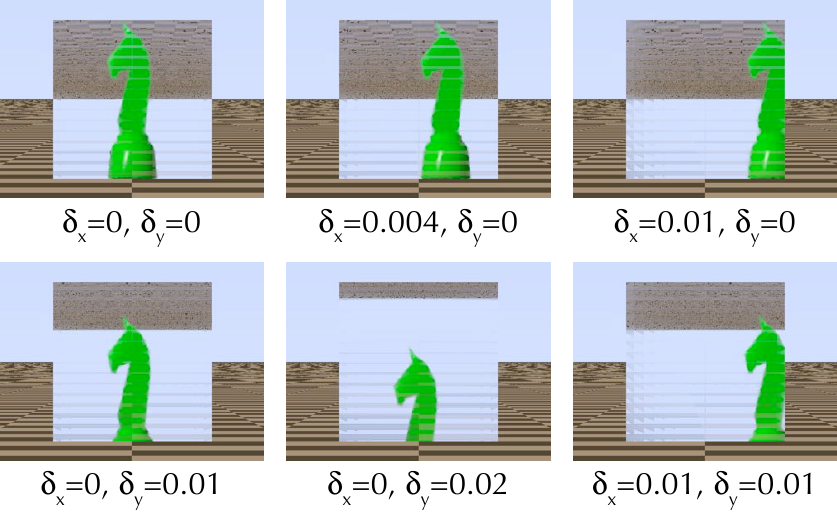}
\caption{\label{vary-delta-figure}
Examples of simulated views through CLAs with different combinations of $\delta_x$ and $\delta_y$.
The views were calculated with the same parameters as those used in Fig.\ \ref{vary-etaX-figure}, but with $\eta_x = -1/4$. 
}
\end{figure}

As the second step of the generalisation we offset the optical axis of the second lens with respect to that of the first.
(The optical axes of the two lenses remain parallel.)
The offset $d$ in the two-dimensional case generalizes into offsets $d_x$ and $d_y$ in the $x$ and $y$ direction, respectively.
Again, we describe these offsets by dimensionless quantities
\begin{equation}
\delta_x = \frac{d_x}{f_{x1}}, \quad \delta_y = \frac{d_y}{f_{y1}}.
\end{equation}
Fig.\ \ref{vary-delta-figure} shows views through CLAs with different combinations of $\delta_x$ and $\delta_y$.

As the final generalisation step we rotate the combination of two elliptical lenses.
We describe this rotation by Euler angles $(\phi, \theta, \psi)$, using the most common definition of the Euler angles (the $\chi$-convention) \cite{Weisstein-2008-Euler-angles}, in which the general rotation is performed in the form of three rotations:  first by an angle $\phi$ about the $z$ axis, then by an angle $\theta$ about the $x$ axis, and finally by an angle $\psi$ about the $z$ axis again.
In the following we briefly discuss each of these angles, and examples of the effect they have on the view through CLAs.

\begin{figure}
\centering \includegraphics{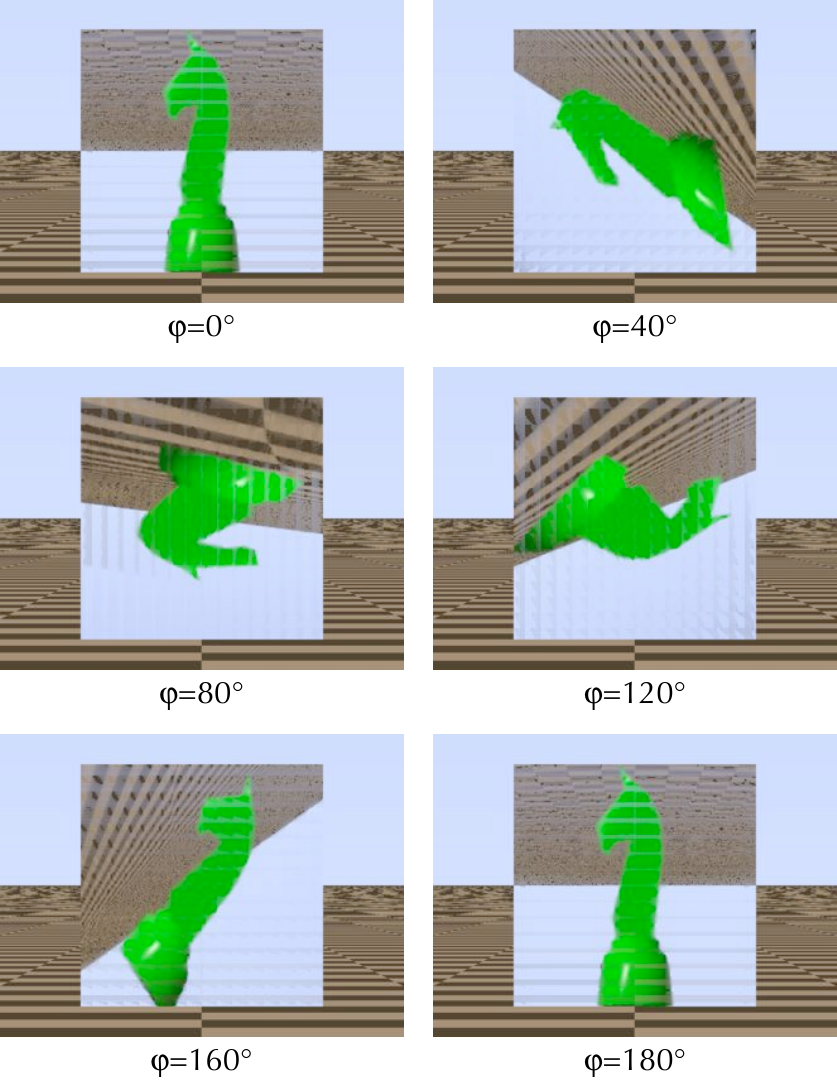}
\caption{\label{vary-phi-figure}
Examples of the effect of a change in the angle $\phi$ on the view through CLAs.
The frames are taken from a movie showing the view through the CLAs as its angle $\phi$ is increased from $0^\circ$ to $180^\circ$.
The movie is available in the supporting online material (MPEG-4, 320kB).
The views were simulated for the same parameters as in Fig.\ \ref{vary-etaX-figure}, but with $\eta_x=-1/4$.
(The optical-axis offsets are zero, i.e.\ $\delta_x=0$ and $\delta_y=0$, and so are the other angles, i.e.\ $\theta = 0$, $\psi = 0$.)}
\end{figure}

The angle $\phi$ rotates the major and minor axes of the elliptical lenses away from the $x$ and $y$ directions, and therefore provides the generalisation mentioned above.
Fig.\ \ref{vary-phi-figure} demonstrates the effect of increasing $\phi$ from $0^\circ$ to $180^\circ$.
The effect on the appearance of the chess piece a finite distance behind the CLAs is a distortion and a rotation through $360^\circ$.
In contrast, the effect on the appearance of the horizon is not a full rotation, but a seesawing, as can be seen from the movie in the supporting online material (see caption of Fig.\ \ref{vary-phi-figure}).

\begin{figure}
\centering \includegraphics{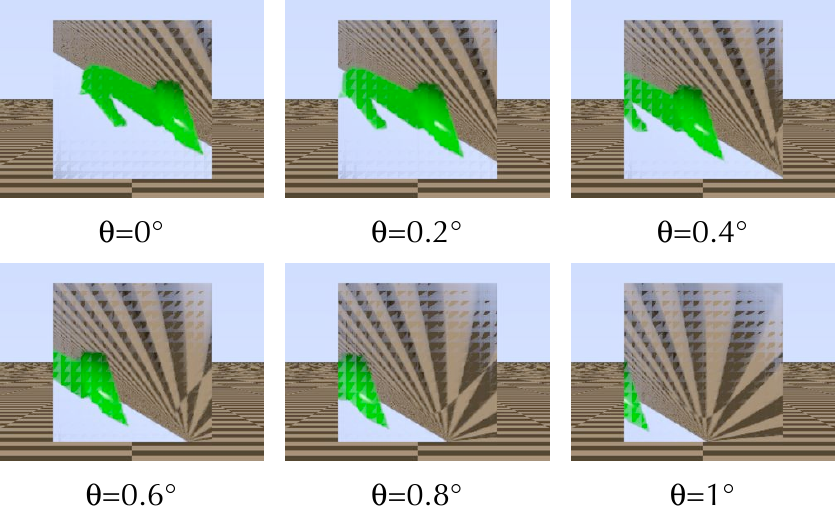}
\caption{\label{vary-theta-figure}
Examples of the effect of changes in the angle $\theta$ on the view through CLAs.
The views shown here were simulated with the parameters used in Fig.\ \ref{vary-phi-figure},
but with $\phi = 45^\circ$ and with different values of $\theta$ as indicated below each frame.}
\end{figure}

The angle $\theta$ describes the inclination of the lenses relative to the planes of the corresponding lenslet arrays.
It is the 3D generalisation of the angle of the same name we discussed in the 2D case (section \ref{2D-section}, particularly Fig.\ \ref{lens-figure}).
Fig.\ \ref{vary-theta-figure} shows example views through CLAs that differ only in their value of $\theta$.
Changing $\theta$ shifts the apparent position of the chess piece seen through the CLAs in each frame sideways.
This effect is very similar to that of changing $\delta_x$ and $\delta_y$; specifically, in the case $\phi = 0$, changing $\theta$ is approximately the same as changing $\delta_x$.
This can be seen very clearly in the 2D case: if equation (\ref{tan-equation}) is written for small angles $\alpha_1$, $\alpha_2$ and $\theta$, it becomes
\begin{equation}
\alpha_1-\theta \approx \delta + \eta (\alpha_2-\theta),
\end{equation}
which can be written as
\begin{equation}
\alpha_2 \approx \frac{\alpha_1}{\eta} + \frac{(\eta-1) \theta - \delta }{\eta}.
\end{equation}
For small angles, $\delta$ and $\theta$ are therefore not independent but simply act as a combined angle that gets added to the scaled angle of incidence, $\alpha_1 / \eta$, to give the angle of refraction, $\alpha_2$.

\begin{figure}
\centering \includegraphics{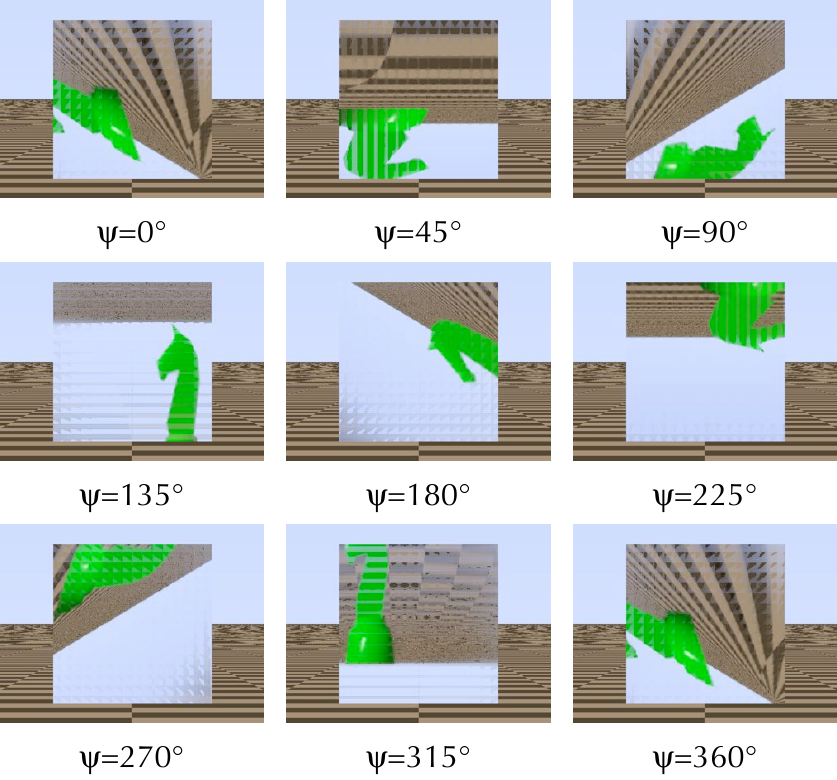}
\caption{\label{vary-psi-figure}
View through CLAs with various values of $\psi$.
All views were simulated for the same choice of parameters as that used in Fig.\ \ref{vary-theta-figure}, and with $\theta = 0.5^\circ$.
The frames are taken from a movie showing the view through CLAs as its value of $\psi$ is increased from $0^\circ$ to $360^\circ$.
The movie is contained in the supporting online material (MPEG-4, 556kB).
}
\end{figure}

\begin{figure}
\centering \includegraphics{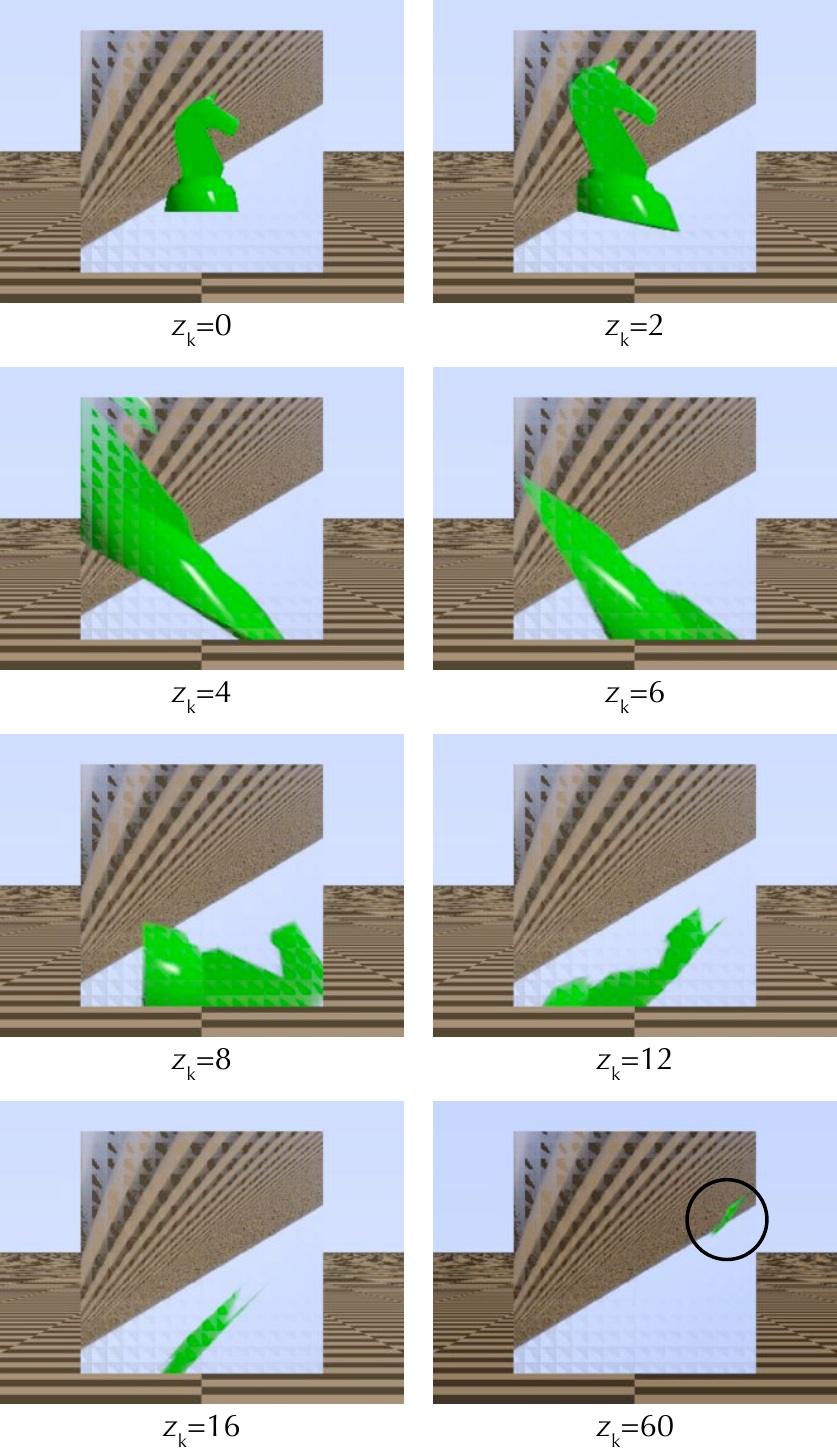}
\caption{\label{vary-knightZ-figure-2}
Effect of changing the distance between the CLAs and the object, calculated for more complex CLAs than that for which Fig.\ \ref{vary-knightZ-figure} was calculated.
Specifically, the object is a chess piece and its distance behind the CLAs, $z_k$, is indicated below each frame.
In the final frame ($z_k = 60$), the position of the chess piece (which appears very small) is marked by a circle.
The parameters of the CLAs are those of the sheets for which Fig.\ \ref{vary-psi-figure} was calculated, with the additional choice $\psi = 90^\circ$.
A movie showing the view as $z_k$ is increased from $-2$ to $+38$ is contained in the supporting online material (MPEG-4, 216kB).
The ray-tracing image of CLAs shown in Fig.\ \ref{lens-figure}(b) was calculated for the same values of $\eta_x$, $\eta_y$, $\delta_x$, $\delta_y$, $\phi$, $\theta$, and $\psi$, but for a significantly reduced separation between the lenslet arrays and correspondingly significantly thicker lenses.
}
\end{figure}

The angle $\psi$ is required to make the rotation general.
Fig.\ \ref{vary-psi-figure} illustrates the effect of changing this angle.
As was already demonstrated in Fig.\ \ref{vary-knightZ-figure}, the view depends on the distance between sheet and object.
Fig.\ \ref{vary-knightZ-figure-2} demonstrates this again, this time for more complex CLAs.

\section{\label{conclusions-section}Conclusions} 

In this paper we have described generalisations of CLAs \cite{Courtial-2008a} that retain their important property of acting like a homogeneous light-bending sheet.
We have not yet investigated in detail the optics of such generalized CLAs, for example their imaging properties, but intend to do this in future.

It is also possible to use the additional parameters locally describing CLAs as additional degrees of freedom in general optical-design applications.
For example, we believe it will be possible to realize optical cloaking \cite{Leonhardt-2006,Pendry-et-al-2006} from a range of directions with structures consisting of CLAs.

Finally, we intend to build generalized CLAs.

\ack
ACH acknowledges funding by the UK's \emph{Engineering and Physical Sciences Research Council} (EPSRC).
JC acknowledges funding by the Royal Society (London).

\section*{References}


\end{document}